\documentclass[prl, twocolumn, byrevtex, showpacs]{revtex4}
\usepackage{graphicx, wrapfig}
\begin{document}

\title{On the possibility of invoking wave-particle dualism by addressing localized deformation}

\author{Lev B. Zuev}

\affiliation{Institute of Strength Physics and Materials Science
\\ Russian Academy of Sciences, Siberian Branch, 634021, Tomsk, Russia}
\email[E-mail: ]{levzuev@mail.tomsknet.ru}
\pacs{05.45.Df, 05.70.Ln, 62.29.Fe.}

\begin{abstract}

The deBroglie relation is applied to the analysis	of auto-wave processes of localized plastic flow in various materials and the results obtained are considered. It is found that the localization of plastic deformation can be conveniently addressed by invoking a hypothetical quasi-particle conjugated with the wave process of flow localization. The mass of the quasiparticle and the area of its localization have been defined.

\end{abstract}

\maketitle

It has been found experimentally (see, for example,~\cite{ref1, ref2, ref3}) that over the entire course of flow in solids, plastic deformation is prone to localization.  The form of localization is determined by the law of work hardening acting at a given stage of flow; hence a variety of local strain patterns observed.  The most striking type of localization is observed during linear work hardening of single- and polycrystals with strain-independent work hardening coefficient, i.e. $\theta=\frac{1}{G}\frac{d\tau}{d\epsilon}$  ($G$ is the shear modulus; $\tau$ and $\epsilon$ are the shear stress and the strain, respectively).  In this case, in the deforming specimen there emerges a typical wave picture, for which the wavelength $\lambda$ and the wave propagation velocity, $V$, which is inversely proportional to $\theta$, can be measured~\cite{ref3}.  The observed waves are generated as a result of interaction and self-organization of elementary acts of plastic deformation; therefore, they can be assigned~\cite{ref3} to a certain kind of dissipative structures that form in open systems~\cite{ref4}, to which deforming solids also belong.  

By addressing this kind of waves in~\cite{ref5}, the author applied the de Broglie relation to calculate mass.  It was found that the calculated value correlated with the atomic weight of the metal from which the specimen was made.  However, this result was obtained by assuming $V$ to be equal to the motion velocity of the movable clamp of the test machine, i.e. $V =V_{mach}$ , which was apparently an unjustified assumption, since the experimental evidence reported in~\cite{ref2, ref3} suggests that $10 \leq V/V_{mach} \leq 50$.  Therefore, in the present work an attempt is made to use the de Broglie relation in a more consistent fashion for the treatment of experimental data on the wavelength of localized deformation, $\lambda$, and the rate of wave propagation, $V$, which were obtained previously~\cite{ref1, ref2, ref3} for Cu, $\gamma$-Fe and Ni single crystals and Zr and Al polycrystals.  

Indeed, using experimental values of $\lambda$ and $V$~\cite{ref3}  and the de Broglie relation $\lambda=\frac{h}{mV}$ ($h$ is the Planck constant) in the same way as it was done in~\cite{ref5} for estimating mass
\begin{equation}
m=\frac{h}{\lambda V}
\label{EqPlanck}
\end{equation}
$m$ values were calculated for five metals investigated.  The estimates obtained are listed in the Table.  Evidently, the listed values fall within a narrow range; in all of the cases, $m_e \ll m \approx 1$ amu = 1.66$\cdot 10^{-24}$g ($m_e$ is the rest electron mass; amu is the atomic mass unit) and the average atomic mass obtained for five metals investigated, $\langle m\rangle= (2.35 \pm 0.70)\cdot 10^{-24}$ g = 1.43 $\pm$ 0.42 amu.

Then division of $m$ by the density $\rho$ of the respective metal yields volume $\Omega$, i.e.
\begin{equation}
\Omega=\frac{m}{\rho}
\label{eqOmega}
\end{equation}  

The values of $\Omega$ and $d_{\Omega}=\sqrt[3]{\Omega}$ are listed in the Table.  It turns out that the $d_{\Omega}$ quantities are close to the ion radii, $r_{ion}$, of  the respective metals (in the case of Zr, Fe and Ni, they are practically the same). The quantities $d_{\Omega}/r_{ion}$ obtained for all the metals investigated (see the Table) support the above conclusion; thus it is seen that the average obtained for five elements (Cu, Al, Zr, Fe and Ni) $\langle d_{\Omega}/r_{ion}\rangle $ = 1.05.  

\begin{figure}[b]

\includegraphics[width=0.5\textwidth]{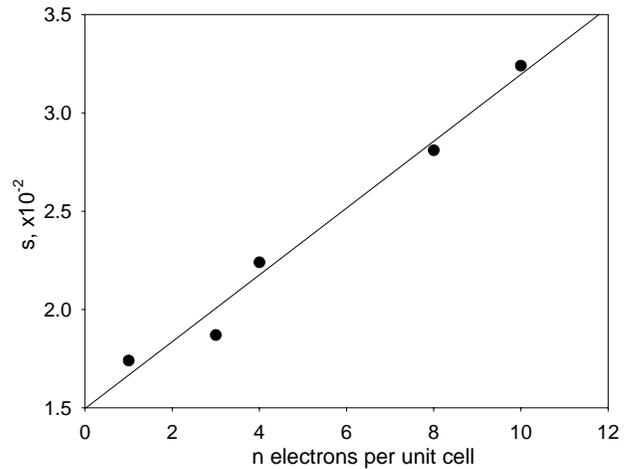}
\caption{Dependence of the parameter $s$ on the number of electrons within unit cell $n$.}
\end{figure}

\begin{table*}[t]
\begin{ruledtabular}

\begin{tabular}{lccccccccc} 
Metal& $\lambda$ & $V, \cdot 10^{-3}$ & $m, \cdot 10^{-24}$ & $\rho$& $\Omega,\cdot 10^{-24}$& $d_{\Omega}$  & $r_{ion}$  & $d_{\Omega}/r_{ion}$ & $s, \cdot 10^{-2}$ \\ 
(number of electrons &(cm)  & (cm/s)  & (g)&(g/cm$^3$ ) &(cm$^3$)  & (\AA)& (\AA)& & \\
 per unit sell, $n$) & & &(amu)  & & & & & & \\ \hline
Cu (1) & 0.45 & 8.0 & 1.84 & 8.9 & 0.21 & 0.59 & 0.72 & 0.82 & 1.74 \\
 & & & (1.10) & & & & & & \\
Al (3) & 0.72 & 11 & 0.84 & 2.7 & 0.31 & 0.68 & 0.51 & 1.33  & 1.87 \\
 & & & (0.50) & & & & & & \\
Zr (4) & 0.55 & 3.5 & 3.44 & 6.5 & 0.53 & 0.81 & 0.79 & 1.02 & 2.24 \\
 & & & (2.07) & & & & & & \\
Fe (8) & 0.50 & 5.1 & 2.60 & 7.9 & 0.33 & 0.69 & 0.64 & 1.08 & 2.81 \\
& & & (1.57) & & & & & & \\
Ni (10) & 0.35 & 6.0 & 3.16 & 9.9 & 0.32 & 0.68 & 0.69 & 0.99 & 3.24 \\
 & & &(1.90)  & & & & & & \\
\end{tabular}
\end{ruledtabular}
\caption{Estimates of the macro-characteristics of the wave processes}
\label{Table}

\end{table*}

Let us consider a certain tendency revealed by the series of $m$ values derived from~(\ref{EqPlanck}) for the metals investigated.  It turns out that the quantity $s$, which was obtained for the corresponding elements by normalization of $m$ values with respect to atomic mass, $M_{at}$, 
\begin{equation}
s=m/M_{at} \ll 1
\label{EqS}
\end{equation}
grows linearly with the number of electrons within unit cell, $n$, of the respective metal~\cite{ref6} in the range covered by the metals investigated, i.e. $1\leq  n \leq 10$.  In this instance, the equation $s(n)$ has the form (see Figure)

\begin{equation}
s=s_0+\kappa n=1.5\cdot 10^{-2}+0.17 \cdot 10^{-2} \cdot n
\label{EqSN}
\end{equation}                                        
and the correlation coefficient for $s$ and $n$ has a rather meaningful value of 0.99~\cite{ref7}.

Finally, consider another possibility of mass evaluation, which is associated with localization waves of plastic flow.  It is shown in~\cite{ref3} that auto-waves of localized plastic deformation are governed by dispersion law of the type $\omega=1+k^2$, which is tantamount to the existence of a gap in the vibration spectrum $0 \leq \omega \leq \omega_{min}\approx 10^{-2}$s$^{-1}$.  Accordingly, by making use of the averaged value of wave velocity derived for five metals investigated, $ \langle V \rangle = (6.72 \pm 1.29)\cdot 10^{-3}$ cm/s, one can calculate average mass
\begin{equation}
 \langle m \rangle=\frac{2\hbar \omega_{min}}{\langle V \rangle ^2}\approx (2.93\pm 0.49)\cdot 10^{-24}g \approx (1.8\pm 0.3) amu
\label{EqGap}
\end{equation}which has the same order of magnitude as the above value  .  Moreover, a comparison of the both averaged values~\cite{ref7} reveals that the difference in the estimates obtained by the above two methods is statistically insignificant.  It is also obvious that  $\hbar \omega \ll  k_BT$( $k_B$ is the Boltzmann constant), which suggests that spontaneous excitation of auto-waves of localized deformation is liable to occur at any temperature.  
Although straightforward, the above estimates are in no way trivial.  They make it apparent that to the macroscopic characteristics of the process of plastic flow localization, such as the propagation rate and wavelength of the wave of localized deformation, directly correspond microscopic entities with physically meaningful parameters, such as size, $\sim r_{ion}$, and mass $\sim$ 1 amu.  In point of fact, we are dealing with a correspondence between the collective processes of self-organization (auto-waves) occurring in the deforming medium on the one hand and certain quasi-particles having:
\renewcommand{\theenumi}{\roman{enumi}}

\begin{enumerate}

\item effective mass, $0.5 \leq m \leq 2$ amu, 
\item  size (area of localization),  $d_{\Omega}\approx r_{ion}$ , 
\item	 velocity, $10^{-3} \leq V \leq 10^{-2}$ cm/s
\end{enumerate} 
on the other.

Such quasi-particles might be correlated with the auto-wave processes of self-organization occurring in the course of plastic flow and with the propagation of auto-waves having the above macroscopic characteristics.


\begin{thebibliography}{8}

\bibitem{ref1} L.B. Zuev, V.I. Danilov, N.V. Kartashova, JETP Lett. 60, 553 (1994).
\bibitem{ref2} L.B. Zuev, V.I. Danilov, Phil. Mag. 79, 43 (1999). 
\bibitem{ref3} L.B. Zuev, Ann. Phys. 10, 965 (2001).
\bibitem{ref4} G. Nicolis, I. Prigogine, Exploring complexity. New York: W.H. Freeman, 1989.
\bibitem{ref5} J.P. Billingsley, Int. J. Solids Structures 38, 4221 (2001).
\bibitem {ref6}	A.P. Cracknell, K.C. Wong, The Fermi surface. Its concept, determination, and use in the physics of solids. Oxford: Clarendon Press, 1973.
\bibitem{ref7} D. Hudson, Statistics. Geneva: CERN, 1964. 
\end{thebibliography}
\end{document}